# Linearly polarized, Q-switched Er-doped fiber laser based on reduced graphene oxide saturable absorber

Grzegorz Sobon[1], Jaroslaw Sotor[1], Joanna Jagiello[2], Rafal Kozinski[2], Krzysztof Librant[2], Mariusz Zdrojek[3], Ludwika Lipinska[2], and Krzysztof M. Abramski[1]

[1] Laser & Fiber Electronics Group, Wroclaw University of Technology, Wybrzeze Wyspianskiego 27, 50-370 Wroclaw, Poland
[2] Institute of Electronic Materials Technology, Wolczynska 133, 01-919 Warsaw, Poland
[3] Faculty of Physics, Warsaw University of Technology, Koszykowa 75, 00-662 Warsaw, Poland

We demonstrate generation of linearly polarized pulses from a passively Q-switched Erbium-doped fiber laser. The cavity was designed using only polarization maintaining (PM) fibers and components, resulting in linearly polarized output beam with degree of polarization (DOP) at the level of 97.6%. Reduced graphene oxide (rGO) was used as a saturable absorber for Q-switched operation. The laser was capable of delivering 1.85 µs pulses with 125 nJ pulse energy at 115 kHz repetition rate.

Graphene, due to its unique nonlinear optical properties, has been widely used as a saturable absorber (SA) for various types of lasers. Since the demonstration of a mode-locked fiber laser [1], many setups were presented, utilizing erbium- [1-4], ytterbium- [5,6] and thulium-doped fibers [7]. Thanks to the high modulation depth, broad operation range and ultrafast recovery time, graphene may be used to generate ultrashort pulses (e.g. 174 fs [4]) from passively mode-locked lasers. Recently, also passive Q-switching with graphene-SA has been demonstrated [8-11]. In a passively Q-switched laser, the cavity losses are automatically modulated by the saturable absorber, so there is no need of placing an active component (e.g. acousto-optic modulator) inside the resonator. The repetition rates of such lasers are lower and the pulses are longer in comparison to mode-locked lasers, but the achieved pulse energies usually exceed those which are achievable from mode-locked oscillators. Recent reports have shown the possibility of achieving 40 nJ pulses from graphene Q-switched Er-doped lasers [8]. Nevertheless, all previous reports on graphene Q-switched lasers were related to non-polarization maintaining cavities. Therefore, the polarization state of the output beam was undetermined. Such setups require polarization controllers (PC) placed in the laser cavity in order to induce the pulsed operation. In many practical applications (industrial or scientific) it is extremely important to maintain the linear polarization state of the laser output. In general, the saturable absorption in graphene is a polarization-independent process, which was verified experimentally by Zhang et al. [12]. It means, that graphene-based saturable absorbers may be used in PM cavities with only one, well-defined polarization state, without affecting the saturable absorption. This opens a possibility to generate scalar solitons in mode-locked lasers [13,14] or linearly polarized pulses from Q-switched lasers, with the DOP near to 100%.

Up till now, many various methods were used to fabricate graphene saturable absorbers. For example, Bao et. al and Zhang et. al [1,2] demonstrated the usage of chemical vapor deposition (CVD) to synthetize graphene on Ni films. Also mechanical exfoliation of graphene from bulk graphite may lead to efficient fiber laser mode-locking [15]. Wet chemistry offers a wide spectrum of techniques of preparing flake graphene and its derivatives. Among them two methods are most attractive and efficient. The first is direct exfoliation of graphite in organic solvents (e.g. dimethyloformamide (DMF) [16]) or in aqueous solutions with surfactant with the use of ultrasound power. Such obtained graphene flakes suspension also may be applied to optical substrates or fiber connectors, even if mixed with polyvinyl alcohol (PVA) [3,4]. The second method appropriate for preparing saturable absorbers based on graphene oxide (GO) is an oxidation/reduction process [17,18]. GO is an atomically thin sheet of carbon covalently bonded with oxygen functional groups, what damages the structure of $sp^2$ hybridized carbon atoms. It may be reduced using various agents, like sodium borohydride [19], to repair defects and create more $sp^2$ domains. Also, the polymer composites with reduced GO, i.e. polyvinylidene fluoride (PVDF)/rGO are known as saturable absorbers [20].

In fully fiberized lasers, controllable covering of a connector tip by graphene flakes dispersed in solutions may be challenging. For this purpose most often the optically driven deposition is used [16,21]. This method was previously applied to solutions based on carbon nanotubes [22]. Although, it is very challenging to optimize the parameters of the process (laser wavelength, power, duration) to obtain desired graphene layer thickness.

In this work, we demonstrate a fully polarization maintaining, passively Q-switched Erbium-doped fiber laser, based on rGO saturable absorber. The designed cavity incorporates only PM fibers and components, which allows to generate pulses with well-defined, linear polarization state, making such source suitable for many industrial and scientific applications. We have applied a very simple and reproducible method of deposition of graphene layer onto the fiber, using small droplet of graphene solution dropped to the end-face of one connector by a micropipette and subsequently dried. The number of graphene flakes can be controlled by the solutions concentration and volume of the droplet. This simple method of preparing fibers with graphene saturable absorbers is cost effective and very attractive for mass production.

* Corresponding author: Grzegorz Sobon, e-mail: grzegorz.sobon@pwr.wroc.pl, Phone: +48713445689, Fax: +48713203189.

Graphene oxide was prepared using a modified Hummers method from expanded acid washed graphite flakes [23,24]. Briefly, expanded graphite was immersed in concentrated $H_2SO_4$ and $NaNO_3$ mixture. Next, $KMnO_4$ was slowly added and it was left overnight. To finish the oxidation process, deionized water and concentrated $H_2O_2$ were added. Post reaction mixture was purified using hydrochloric acid solution and deionized water. Reduction of GO was carried out by a simple, quick and cost effective method [25,26]. 5 ml of GO water suspension (concentration 10 mg/ml) was diluted with 200 ml of deionized water and 10 ml of 57% concentrated HI was added. Reduction was carried for 2 h in $80^0C$ with stirring, which could be easily observed - GO changed color from brown to black and it precipitated from suspension. After that, rGO was filtered and washed with deionized water. Reduced graphene oxide was suspended in 50 ml of water and 0.05 g of sodium deoxycholate (SDC) was added. The mixture was stirred until salt dissolved and then it was sonificated for 30 minutes. After this ultrasound-assisted functionalization by SDC, rGO formed a stable suspension. As prepared liquid was transferred on the angle-polished fiber connector (FC/APC) end using micropipette. The fiber was left to dry on air for 2 h.

In order to check the preparation of the graphene flakes we performed Raman spectroscopy of as-prepared rGO and GO samples. The spectra were recorded by Dilor XY-800 spectrometer, using 514 nm beam of an argon-ion laser. As it is shown in Fig. 1, there are two main peaks at around 1350 and 1600 $cm^{-1}$, the D and the G, respectively, whereas Raman spectrum for pristine graphite/graphene should contain G peak only [25]. The G peak originates from in-plane vibration of $sp^2$ carbon atoms, it is due to the $E_{2g}$ phonon at the Brillouin zone center [27,28]. The D peak indicates the presence of defects and disorder in the graphite/graphene structure. This mode arising from the doubly resonant disorder-induced mode. In this case, this is the result of oxygenation of graphite to graphene oxide. Such structure contains covalently bonded oxygen functional groups, e.g. hydroxyl, epoxy, carbonyl, what destroy the network of conjugated double bonds. The D peak originates from the breathing modes of six-membered rings that are activated by defects [27]. After the reduction the $I_D/I_G$ intensity ratio increases from 1.21 for GO (inset Fig. 1) to 1.45 for rGO. This suggests, that new or more graphitic domains are formed during the reduction and leads to the conclusion, that GO has been significantly reduced. The atomic ratio of oxygen measured by X-ray photoelectron spectroscopy (XPS) is about 33 % for GO (and 67% of C atoms) and less than 9% for rGO (and 91% of C atoms). Thus about 73% of oxygen present in the structure of GO was removed after the reduction process. So far there are no reports in the literature on the total elimination of oxygen functional groups. It is very difficult to achieve due to large variety of these groups, their unclear distribution in the GO lattice and interactions.

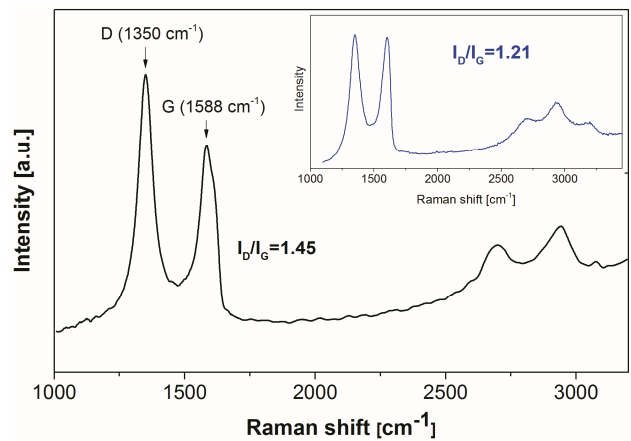

*Fig. 1. Raman spectra of rGO and GO (inset) recorded with 1 mW of 514 nm wavelength laser line.*

The experimental setup of the Q-switched laser is shown in Fig. 2. The resonator consists of a 50 cm long PM erbium-doped fiber (Nufern PM-ESF-7/125), a fiber isolator, 980/1550 filter-type wavelength division multiplexing (WDM) coupler, 30% output coupler and the rGO saturable absorber. The laser is counter-directionally pumped by a 980 nm laser diode with maximum available power of 175 mW. All used fibers and components were polarization maintaining (Panda-type) and were spliced using a Fujikura FSM-45PM fusion splicer with θ-axis rotation (suitable for PM components), in order to match the appropriate polarization axis. The total cavity length is around 4.2 m. The setup is simplified in comparison to typical, non-PM ring cavities [8,10,21], since it does not require a polarization controller to start the pulsed operation. The laser performance was observed using optical spectrum analyzer (ANDO 6317B), 350 MHz digital oscilloscope (Hameg HMO3524) coupled with a 30 GHz photodetector (OptiLab PD-30) and a polarimeter (Thorlabs PAX5710).

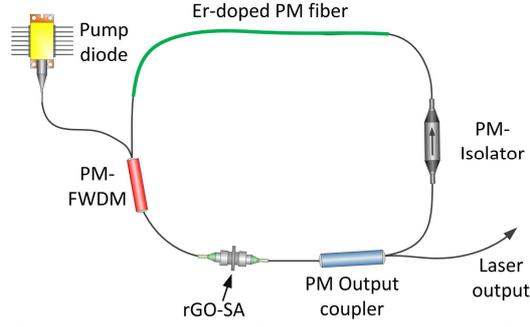

*Fig. 2. Setup of the Q-switched fiber laser. PM-FWDM – polarization maintaining filter-WDM coupler, rGO-SA – reduced graphene oxide saturable absorber.*

After launching the pump at low power level (below 120 mW) the laser operates in continuous-wave (CW) regime. Above the 120 mW threshold, Q-switched operation is observed. Figure 3 shows the optical spectrum of the laser, recorded at the maximum available pump power. The multi-peak structure of the optical spectrum might be caused by the cavity birefringence, which favors generation of several longitudinal modes. Such behavior in passively Q-switched fiber lasers was reported previously by Wang et al. [29], where the birefringence of the resonator resulted in dual-wavelength operation.

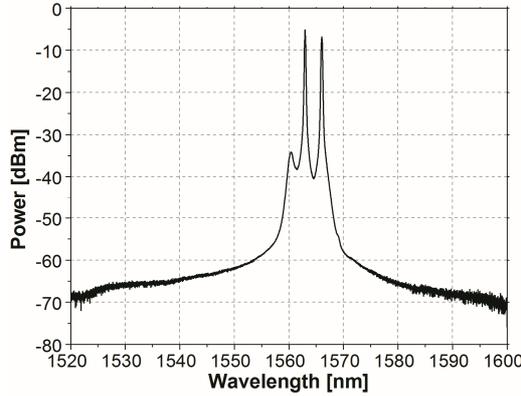

*Fig. 3. Optical spectrum of the laser recorded with 0.05 nm resolution.*

The repetition rate, pulse duration and pulse energy are pump-power dependent and may be tuned only by changing the pump power, which is a typical behavior of passively Q-switched fiber lasers. In our case, the repetition frequency might be tuned from 104 kHz to 116 kHz and the pulse duration from 3.85 µs to 1.85 µs by changing the pump from 120 mW to 175 mW (see Fig. 4). The achieved pulse durations are comparable to previous reports on graphene Q-switched fiber lasers [8,10,21].

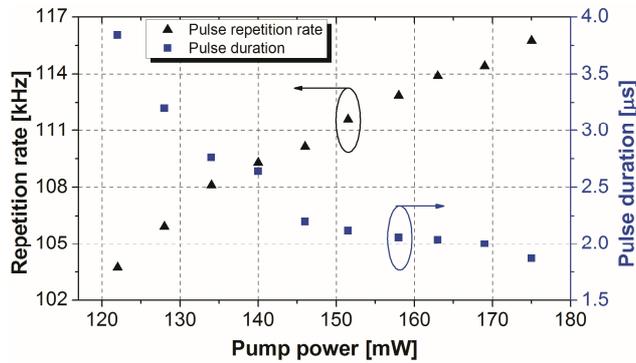

*Fig. 4. Pulse duration and the repetition frequency as a function of the pump power.*

An exemplary pulse train recorded at 115 kHz repetition frequency is plotted in Fig. 5a. The 1.85 µs pulse shape is shown in Fig 5b. Both traces were observed at the maximum pump power (175 mW).

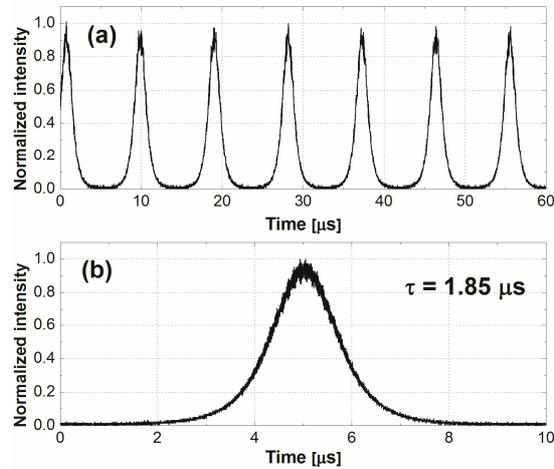

*Fig. 5. Recorded pulse train (a) and a single pulse (b) at 115 kHz repetition rate.*

Depending on the pump power, the output pulse energy was varying between 90 nJ and 125 nJ. In comparison, previous reports showed 33 nJ [9] and 40 nJ [8] with higher pump powers (up to 200 mW). The maximum average output power of the laser was 14.6 mW at 175 mW of pumping. It means, that the efficiency of the laser was at the level of 8.3%, which is few times better than reported previously [8,9]. The dependence of the pulse energy and average output power on the pump power is linear in the whole available pump range (see Fig. 6).

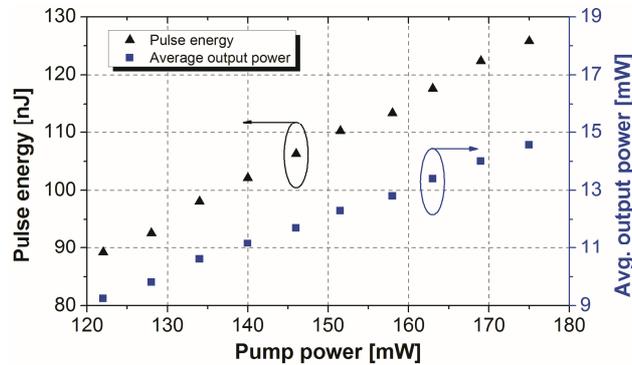

*Fig. 6. Pulse energy and average output power as a function of the pump power.*

In order to characterize the polarization state of the laser output, the output beam was analyzed using a polarimeter. The measurement was performed at the maximum pump power (175 mW). One of the most important parameters of a polarized wave is its degree of polarization, which is calculated as the fraction of the total power that is carried by the polarized component of the wave [30]. The results show, that the average DOP of the output beam form the Q-switched laser was at the level of 97.6% with standard deviation of 0.02% over the 60 seconds measurement time (see Fig. 7). The result is comparable to graphene mode-locked fiber lasers based on PM fibers [12,13].

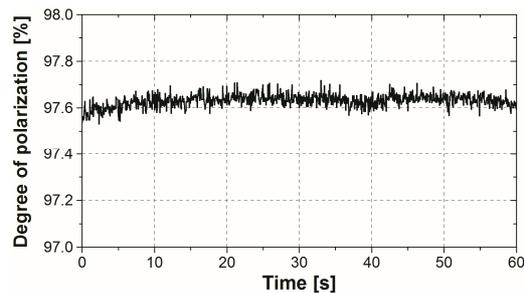
*Fig. 7. Degree of polarization measured over 1 minute period at maximum pump power.*

In conclusion, we have demonstrated a polarization maintaining Er-doped fiber laser, passively Q-switched by a rGO saturable absorber. The SA was fabricated by dropping a small droplet of the prepared rGO solution on the fiber connector end by a micropipette. The laser was capable of generating 1.85 µs pulses with 125 nJ energy at 115 kHz repetition frequency. The basic parameters of the output radiation (pulse duration, energy and repetition rate) might be easily tuned just by changing the pump power. Since the designed cavity incorporates only PM fibers and components, the output beam was linearly polarized with average DOP at the level of 97.6%.

**Acknowledgements**

Work presented in this paper was supported by the National Science Centre (NCN, Poland) under the project "Saturable absorption in atomic-layer graphene for ultrashort pulse generation in fiber lasers" (decision no. DEC-2011/03/B/ST7/00208). Research fellowship of one of the authors (G.S.) is co-financed by the European Union as part of the European Social Fund.